\documentclass[ ]{aa}

\usepackage{graphicx}
\usepackage{color}
\usepackage[varg]{txfonts}
\usepackage{multirow}

\usepackage[round]{natbib}
\begin{document}

\title{Long-period oscillations of active region patterns: least-squares mapping on second-order curves}

\author{G.~Dumbadze\inst{1,3}, B.M.~Shergelashvili\inst{2,3,5}, V.~Kukhianidze\inst{3}, G.~Ramishvili\inst{3}, T.V.~Zaqarashvili\inst{4,3,2}, M.~Khodachenko\inst{2}, E.~Gurgenashvili\inst{3}, S.~Poedts\inst{1} and P.~De Causmaecker\inst{5}
}
\institute{Center for mathematical Plasma Astrophysics, Department of Mathematics, KU Leuven, 200 B, B-3001, Leuven, Belgium\\
                 \and
             Space Research Institute, Austrian Academy of Sciences, Schmiedlstrasse 6, 8042 Graz, Austria\\
                  \and
               Abastumani Astrophysical Observatory at Ilia State University, University St. 2, Tbilisi, Georgia\\
                  \and
             Institute of Physics, IGAM, University of Graz, Universit\"atsplatz 5, 8010 Graz, Austria\\
                  \and
             Combinatorial Optomization and Decision Support, KU Leuven campus Kortrijk, E. Sabbelaan 53, 8500 Kortrijk, Belgium
                                 }
\abstract
{Active regions (ARs) are the main sources of variety in solar dynamic events. Automated detection and identification tools need to be developed for solar features for a deeper understanding of the solar cycle. Of particular interest here are the dynamical properties of the ARs, regardless of their internal structure and sunspot distribution.}
{ We studied the oscillatory dynamics of two ARs: NOAA 11327 and NOAA 11726 using two different methods of pattern recognition.}
{ We developed a novel method of automated AR border detection and compared it to an existing method for the proof-of-concept. The first method uses least-squares fitting on the smallest ellipse enclosing the AR, while the second method applies regression on the convex hull.}
{After  processing the data, we found that the axes and the inclination angle of the ellipse and the convex hull oscillate in time. These oscillations are interpreted as the second harmonic of the standing long-period kink oscillations (with the node at the apex) of the magnetic flux tube connecting the two main sunspots of the ARs.  We also found that the inclination angles oscillate with characteristic periods of 4.9 hours in AR 11726 and 4.6 hours in AR 11327. In addition, we discovered that the lengths of the pattern axes in the ARs oscillate with  similar characteristic periods and these oscillations might be ascribed to standing global flute modes.}
{In both ARs we have estimated the distribution of the phase speed magnitude along the magnetic tubes (along the two main spots) by interpreting the obtained oscillation of the inclination angle as the standing second harmonic kink mode. After comparing the obtained results for fast and slow kink modes, we conclude that both of these modes are good candidates to explain the observed oscillations of the AR inclination angles, as in the high plasma $\beta$ regime the phase speeds of these modes are comparable and on the order of the Alfv\'{e}n speed. Based on the properties of the observed oscillations, we detected the appropriate depth of the sunspot patterns, which coincides with estimations made by helioseismic methods. The latter analysis can be used as a basis for developing a magneto-seismological tool for ARs.}


\titlerunning{Long-period oscillations of active region patterns}

\authorrunning{Dumbadze et al.}

\maketitle

\section{Introduction}
Active regions (ARs) represent  the most complex magnetic structures emerging on the surface of the Sun. They consist of many sunspots. Their number, location, and size vary in time, thus sunspots are indicators or tracers of solar magnetic activity. The complexity of the ARs manifests itself in their morphology and their dynamics. These two features determine the different types of waves and oscillatory motions that abundantly populate the ARs. The study of these oscillatory phenomena can formally be divided into the following branches \citep{Chorley10}: (i) umbral chromospheric oscillations with a typical period of 3 minutes, interpreted as slow magnetoacoustic waves \citep[e.g.,][]{centeno06}. Other interpretation models reported on  acoustic wave propagation in the structured (or stratified) atmosphere \citep{fleckschmits91,kuridze09}; (ii) umbral photospheric oscillations with a typical period of 5 minutes and associated with photospheric acoustic oscillations \citep{thomas84}. In this context, there might be an interesting connection between these oscillations and high-degree p-modes, which themselves are affected by the differential solar rotation \citep{shergelashvili05}; (iii) long-period oscillations \citep[on the order of hours, e.g.,][]{efremov07,goldvarg05}; and (iv) ultra-long-period (torque or torsional) oscillations of sunspot umbrae, with typical periods of several days \citep{khutsishvili98,gopasyuk04}.

In the present paper, we focus on the long-period oscillations in ARs, in particular\ branch (iii) above, that is, those\ with periods of up to a few hours. Observations of these oscillatory features in sunspots have been extensively reported in radio emission measurements by
\citet{gelfreichetal06,efremov07,solovev08,solovev06,Smirnova13}; and \citet{bakunina09}. As has been indicated by \citet{bakunina09}, while the observed short-period oscillations are identified as magnetohydrodynamics (MHD) waves that are trapped inside the magnetic flux tubes of the sunspots, the low-frequency oscillations are caused by quasi-periodic displacements of the whole sunspot as a well localized and stable formation.

It is very important to understand the basic physical nature of the long-period phenomena because they can be directly linked to the excitation of Alfv\'{e}n wave fields in the solar atmosphere. In turn, these wave fields may be engaged in complex conversion and dissipation processes in the solar corona in terms of
processes driven by shear flows \citep[see, e.g.,][]{shergelashvili06}, thermal variability related to wave couplings \citep{shergelashvili07}, and a variety of parametric interactions of waves \citep{zaqrol02,zaqrob02,shergelashvilisw05}.

Automated detection and identification methods for solar magnetic patterns have been developed recently, among them those for ARs and sunspots. The method of AR identification includes morphological analysis and intensity thresholds \citep{Barra09, Preminger01, Zharkov03, Zharkova03, Georgoulis08, Bobra14}. \citet{Zhang10} used Solar and Heliospheric Observatory (SOHO)/Michelson Doppler Imager (MDI) magnetograms to detect regions on the solar surface with a line-of-sight (LOS) component of the magnetic field larger than 250$\;$G and an area more extended than 10$\;$Mm$^2$. \citet{McAteer05} explored and automatically detected ARs on full-disk MDI magnetograms with a threshold value of 50$\;$G used for identification of the AR borders. \citet{Hoeksema14} developed an AR trajectory automatic detection tool, called the Helioseismic and Magnetic Imager (HMI) Active Region Patches (HARPs). It provides primarily spatial information about long-lived coherent magnetic structures at the scale of a solar active region. \citet{martens12} produced software modules that detect, trace, and analyze the emergence and evolution of ARs, magnetic elements and other solar features, such as flares, sigmoids, filaments, coronal dimmings, polarity inversion lines, sunspots and other magnetic structures.

\citet{Verbeeck13} described two algorithms: the solar monitor active region tracker (SMART), which automatically extracts, characterizes, and tracks active regions \citep{Higgins11}; and the automated solar activity prediction (ASAP), which represents a set of algorithms that detect  sunspots, faculae, and active regions \citep{Colak08}. Moreover, \citet{spoca2014} described another method called the spatial possibility clustering algorithm (SPoCA-suite). This method detects ARs, quite Sun (QS), and coronal holes (CH) on full solar disk images. SPoCA-suite segments an extreme ultraviolet (EUV) image into ARs, QSs, and CHs so that every pixel corresponds to one of these three classes, based on their pixel values in the image \citep{Barra09}.

In this paper, we develop an automated AR border detection that is adapted to the study of oscillatory motions. After detecting the boundary points, we build the AR boundary shape using the least-squares fitting to a second-order curve, in particular to an ellipse. Finally, after the data processing and the detection of the ellipse parameter temporal dynamics, we study the possible distribution of the Alfv\'{e}n speed value along the magnetic tubes by interpreting the results as standing kink modes in the thin tube approximation. In our model we use thresholds on the magnetic field gradient instead of its original value, and validate our new concept by comparing it with results obtained with the convex hull method \citep{stenning13}.

This paper is organized as follows: the observations are described in Sect.~\ref{secobservations}. The automated AR border detection and least-squares mapping on the second-order curve of the boundary points are introduced in Sect.~\ref{secmethods}. The results are presented in Sect.~\ref{secresults}. In Sect.~\ref{analysis} we interpret the discovered AR oscillations. Finally, some concluding remarks are presented in Sect.~\ref{secconcl}.

\section{Data and observed ARs}\label{secobservations}
We used data from the SDO/HMI database. This telescope feeds the full-disk solar image in two $4096^2$-pixel CCD cameras that record every 3.75 seconds, giving an overall cadence of 45 seconds for the LOS magnetic field, Doppler and intensity measurements \citep{Schou12, Scherrer12}.
We have selected two active regions, NOAA 11327 and NOAA 11726, which have a sufficiently long lifetime and a topology resembling an elliptical shape. The data set covers the following observational time windows: for AR 11726 -- 20/04/2013 at 00:00 am UT -- 22/04/2013 at 3:12 am UT; for AR 11327 -- 20/10/2011 at 6:00 pm UT  --  22/10/2011 at 9:12 pm UT.

\subsection{AR NOAA 11726}
This AR appeared in the center of the northern hemisphere on 19/04/2013
at about 6:00$\;$am UT and was hidden behind the western limb on 26/04/2013 at 11:00$\;$am UT. On 19/04/2013 the AR emerged from the solar interior as a small round group of sunspots with different polarities. The main spots moved in different directions, and new tiny spots appeared between them. After six hours, another spot with an elongated shape appeared next to the first. On 20/04/2013 at about 2:00$\;$am UT, these two spots joined and formed a single elliptical shape AR (see Fig.~\ref{method} panel$\;$1a). During the whole period of observation, 60 flares were launched from this AR, including 1 M-, 49   C-, and 10$\;$B-class flares. A movie showing the evolution of this AR can be found on the SOLSPANET web site {http://www.solspanet.eu} in the AR catalogs.

\subsection{AR NOAA 11327}
Active region 11327 appeared in the southern hemisphere on 20/10/2011 at about 2:00$\;$am UT and was hidden behind the western limb on 28/10/2011 at 3:00$\;$am UT. The AR emerged on the solar surface as a small elongated bipolar sunspot group. Its main spots were located close to each other and were surrounded by many smaller spos (see Fig.~\ref{method} panel$\;$1b). On 20/10/2011 at 3:00$\;$pm UT, tiny spots filled out the space between the main spots and gave an elliptical shape to the AR. On 22/10/2011 at about 2:00$\;$am UT, the main spots of the AR began to move in opposite directions, which destroyed the elliptical shape. No flares were registered in AR 11327 during this period.

\section{Description of the pattern recognition method (data processing) }\label{secmethods}
\begin{figure}
\includegraphics[width=1.0\linewidth]{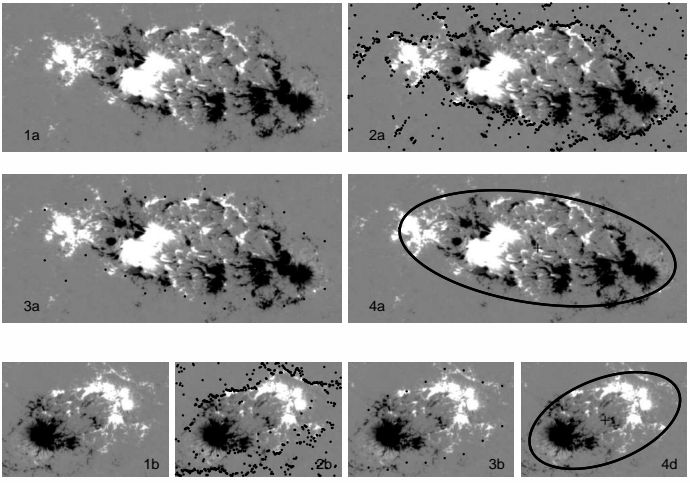}\\
\caption{Snapshots taken on 21/04/2013 at 6:01$\;$am UT for AR 11726 (panels 1a, 2a, 3a, and 4a) and on 21/10/2011 at 12:45$\;$am UT for AR 11327 (panels 1b, 2b, 3b, and 4b). These panels show the stages of our pattern recognition method based on the least-squares mapping of ellipses.}\label{method}
\end{figure}

To study the AR dynamics of interest, we first developed a procedure for the AR  boundary recognition that we call the method of least-squares mapping of an ellipse. This procedure implies several stages, starting from a proper selection of the pixels belonging to the boundary, and followed by the detection of the boundary shape using a least-squares fitting method (see Fig.~\ref{method}).

We considered a rectangular domain enclosing the ARs, which consist of randomly distributed small sunspots and particular pixels with accidently strong magnetic field intensity (see Fig.~\ref{method} panels 1a and b). We identified boundary points by choosing the first pixel with a field gradient exceeding the threshold of 40$\;$Gauss/pixel starting from the top and bottom edges of the domain (see Fig.~\ref{method} panels 2a and b).
We thus detect pairs of $y-$coordinates of the registered pixels in each vertical slice. To reduce the influence of random noise on the distribution of  these points, we took the average of the coordinates for 30 consecutive points and set new number of boundary points (see Fig.~\ref{method} panels 3a and b).
Consequently, we used the found points in the least-squares fitting model (see Fig.~\ref{method} panels 4a and b). The calculation of the ellipse parameters is explained in Appendix~A.

\begin{figure*}
\begin{minipage}[h]{0.5\linewidth}
\center{\includegraphics[width=8.8cm]{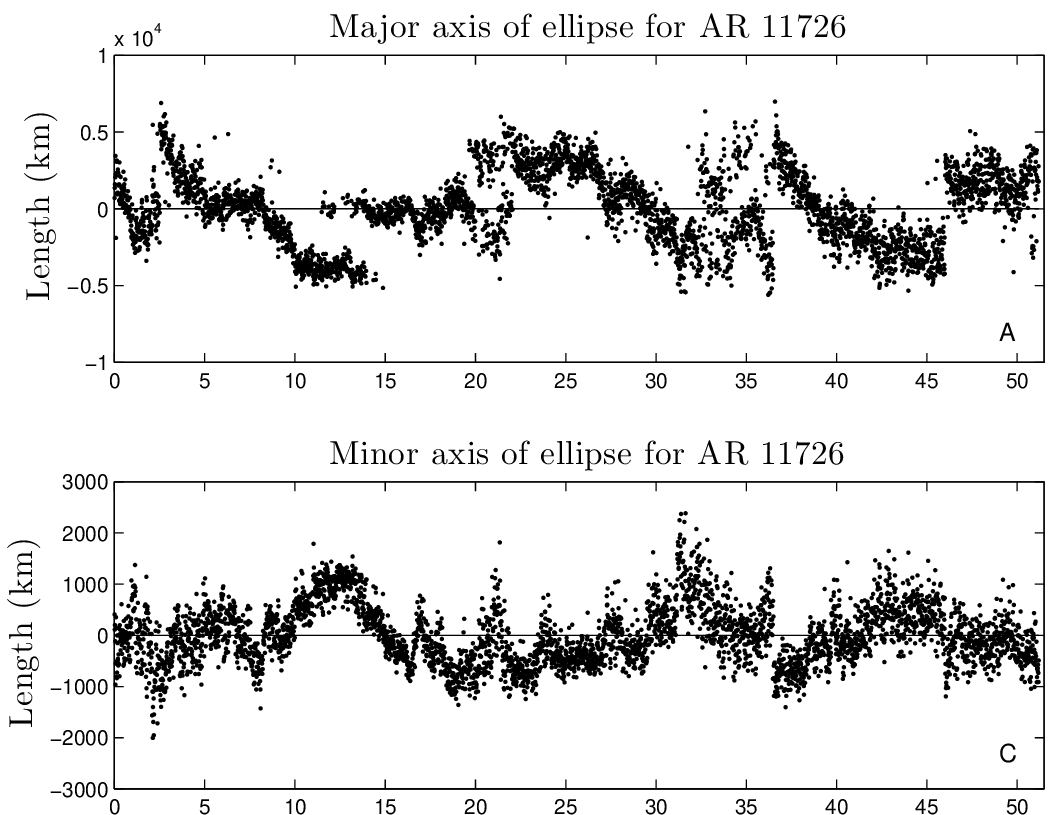}}
\vfill
\center{\includegraphics[width=8.8cm]{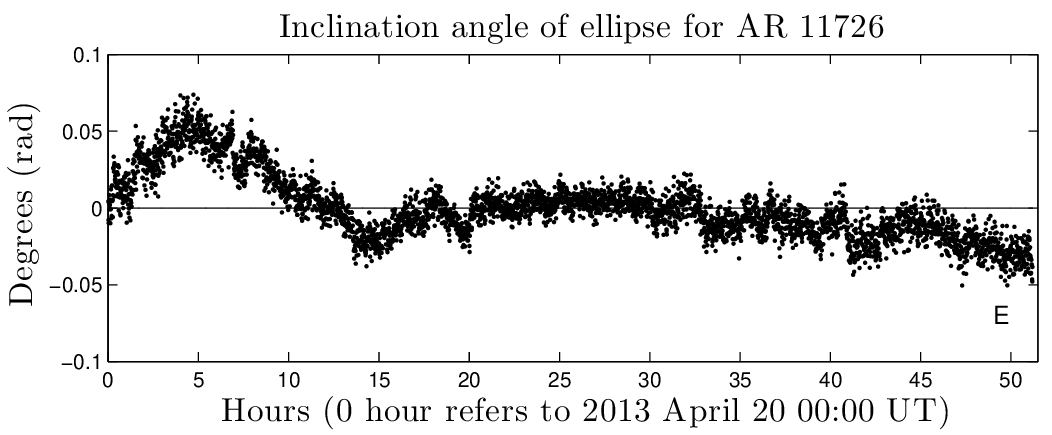}}
\end{minipage}
\begin{minipage}[h]{0.5\linewidth}
\center{\includegraphics[width=8.8cm]{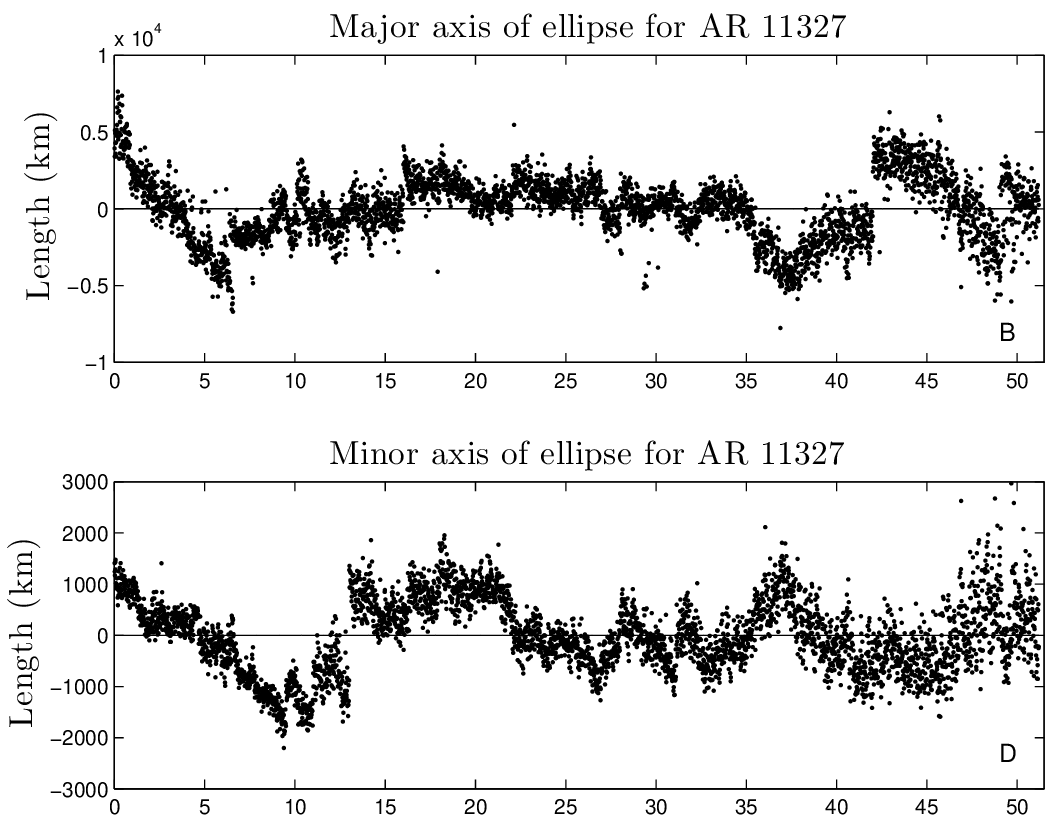}}
\center{\includegraphics[width=8.8cm]{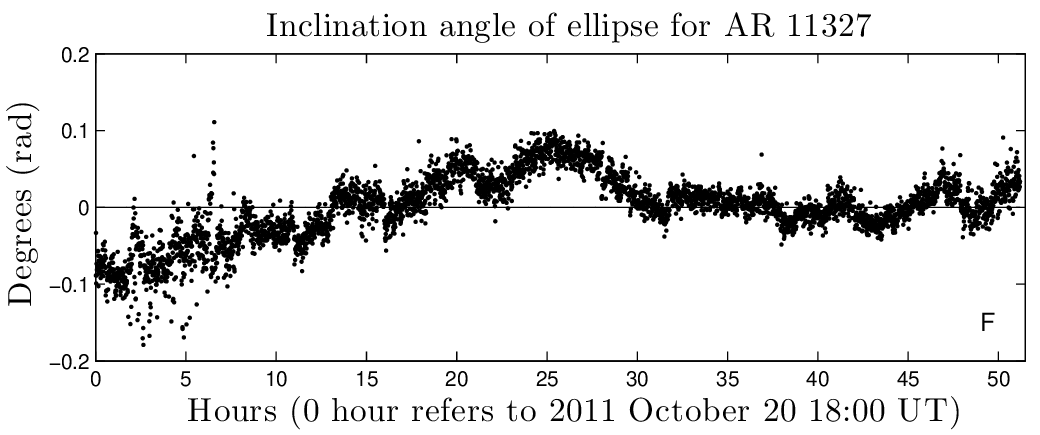}}
\end{minipage}
\caption{The detrended data for the large (top panels), small (middle panels) semi-axis and the inclination angle (bottom panels) for AR 11726 (left panels) and for AR 11327 (right panels), respectively. The vertical axis denotes the length and degree. The horizontal axis indicates the duration of the observations (in hours).
}\label{elldetrend}
\end{figure*}

Using a sequence of snapshots within the observational time span, we produced time series for the major and minor semi-axes and the inclination angle. In general, the axes of both ARs show a tendency of growing in size, approximately linearly. However, the time behavior of their lengths also shows some oscillations, and so does the inclination angle.
The amplitude of the latter varies within the range $[0.09; 0.21]$ (with a mean value of $0.16$) radians for AR 11726; and $[0.26; 0.40]$ (with a mean value of $0.34$) radians for AR 11327.
We conclude that the temporal evolution of the inclination angle oscillates, the major and minor axes grow monotonically, and in addition, the lengths of both axes also oscillate. Therefore, the obtained ellipses  tilt and breathe (as the ellipse axes also oscillate) all the time in a quasi-periodic manner. In Fig.~\ref{elldetrend} we show plots of detrended data by cubic interpolation time series for the axes (top four panels) and the original one for the inclination angle (two bottom panels).

For a reference model as a validation or proof-of-concept of our method, we also employed the convex hull approach to detect the shape of the ARs first and then to study the temporal behavior of the obtained convex hull axis and its inclination angle in time. The convex hull is the smallest convex polygon containing the whole pattern and enables a recognition procedure for the object shape. The boundary points were found by setting a threshold value, and we built the convex hull using the quickhull method \citep{Eddy1977,BYKAT1978}. The quickhull first computes the top, bottom, left, and right points that create a quadrilateral including all subject points and then decreases this quadrilateral till it reaches the boundary points.

Before turning to the detailed description of our results, we list the main advantages of the novel method compared to, for instance, the convex hull method:
\begin{itemize}
  \item Our new method for the registration of potential AR boundary points is based on a threshold for the gradient of the magnetic field instead of its absolute value. This approach allows us to filter out almost all internal pixels and to focus only on the pixels close to the actual AR boundary.
  \item The ellipse shape enables the measurement of both the longitudinal and transverse characteristic scales of the AR, while with the convex hull approach only obtains the former.
  \item Our comparative analysis also showed that the distribution of the inclination angle around its mean value is far more chaotically dispersed with the convex hull method than with our new approach. Hence, we argue that measurements of the inclination angle are much more reliable with the ellipse method.
\end{itemize}

\section{Results}\label{secresults}
\subsection{Ellipse oscillations}
\begin{figure*}[!ht]
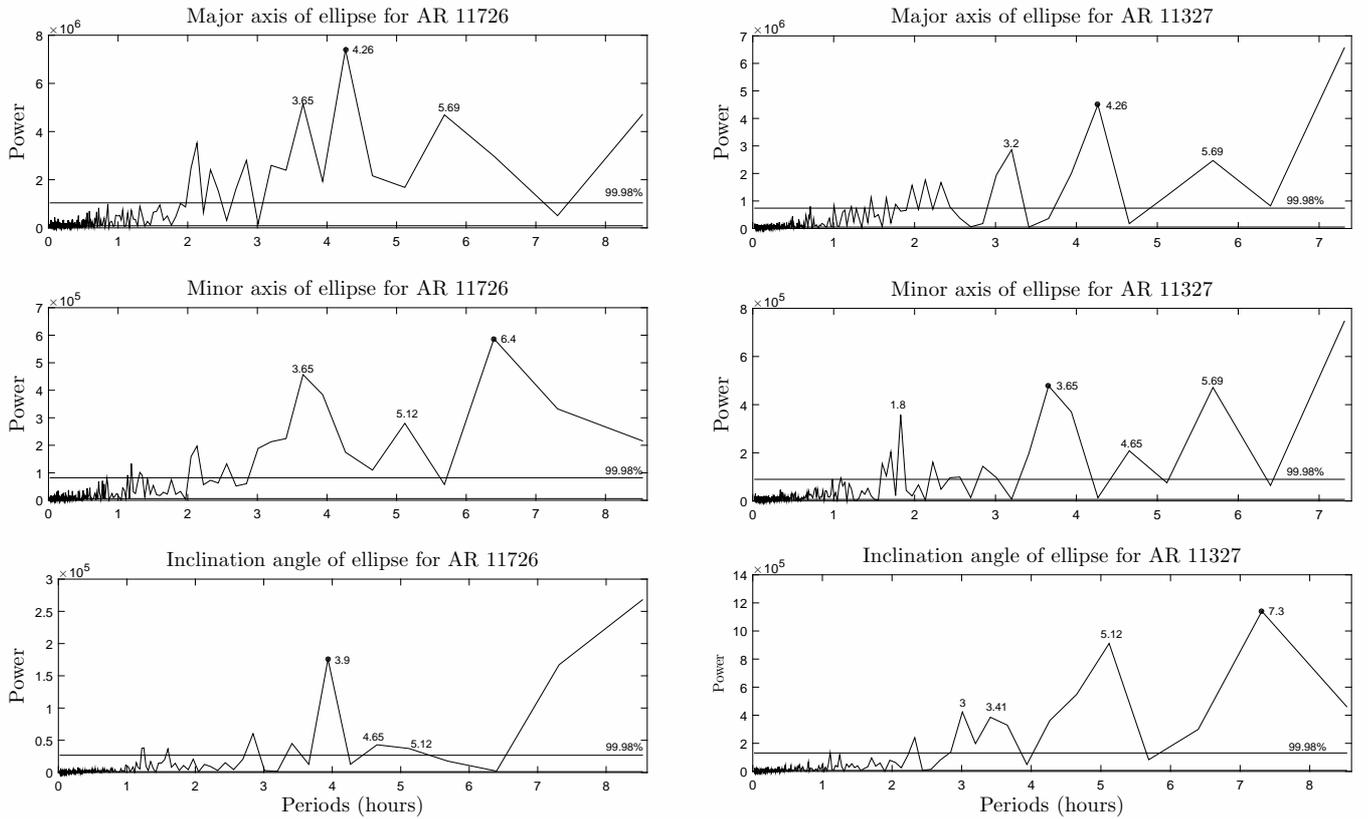

\begin{minipage}[h]{0.5\linewidth}
\center{\includegraphics[width=8.4cm]{fig5a.eps}}
\vfill
\center{\includegraphics[width=8.4cm]{fig5b.eps}}
\end{minipage}
\begin{minipage}[h]{0.5\linewidth}
\center{\includegraphics[width=8.4cm]{fig7a.eps}}
\center{\includegraphics[width=8.4cm]{fig7b.eps}}
\end{minipage}
\caption{Periodicities of the axes lenghts and the inclination angle of the ellipses for AR 11726 (left panels) and for AR 11327 (right panels), respectively. Respectively the vertical axis denotes the power and the horizontal axis denotes the periods of oscillation (in hours). The highest peaks corresponding (or contributing) to the artefact 12 and 24-hour periodicities have been removed.
}\label{ellperiod}
\end{figure*}

To enable a quantitative analysis of the observed oscillations, we applied fast Fourier transform (FFT) techniques to the detrended ellipse axes lengths and to the inclination angle data series. This analysis reveals several significant spectral peaks with a confidence level above 99.98$\%$. The peaks corresponding to the longest periods have been removed because we interpreted them as known instrumental artifacts related to the 12- and 24-hour signal variations in HMI magnetograms \citep{Liu12}. The FFT analysis of the detrended data for axes lengths and for the inclination angle are shown in Fig.~\ref{ellperiod}, the left panels refer to AR 11726 and right panels to AR 11327.

Before we interpret the observed quasi-harmonic behavior, we need to explain the resolution of the obtained Fourier spectra that characterize the periodogram resolution.
First of all, $\Delta f = 1/T$, where $T$ denotes the total duration of the data set. In our case $T\approx 52\;$hours (see Fig.~\ref{elldetrend}), yielding a value of $\Delta f \approx 0.019\;$hour$^{-1}$. The second method is used to calculate the ratio of the sampling frequency and the number of frequency bins in the spectrum $\Delta f=f_s/N_{fft}$. With a 45 second cadence of the observations, we obtain $f_s=80\;$hour$^{-1}$, which gives a similar value $\Delta f\approx 0.019\;$hour$^{-1}$.
We can check the observed peaks of the ellipse data for both ARs as
\begin{equation}\label{resolution12}
    \Delta f_{a,b}=\frac{1}{P_b}-\frac{1}{P_a},
\end{equation}
where $a$ and $b$ are two from any three obtained peaks. The spectral distance between the periods is on the order of the spectral resolution, while to be able to isolate individual peaks well, the distance should be at least a few multiples of the resolution. Therefore, in the current study we do not distinguish between the three peaks, but operate with their average values. The averaging of periods allows us to perform a semi-qualitative or quantitative rough estimation of the periods and related physical parameters of the ARs (see below). This type of estimation and its precision suffices for the current stage of the investigation. A more precise analysis of the spectral peak fine structures requires a more rigorous mathematical approach and is left for future research.

We calculated the characteristic errors of the obtained periods using two different methods. The first method calculates the half-width of the corresponding power peaks and then takes their mean values. The second method is based on finding the standard deviation of the set of peaks from the above derived mean. Next, we estimated the errors by calculating the standard error $\Delta P_{error}=\sigma/\sqrt{n}$ where $n=3$ (ellipse) or $n=4$ (convex hull) is the number of mean period samples. The resulting mean values of the periods and their errors are listed in Table~\ref{table}.

\subsection{Convex hull oscillations}
\begin{figure*}[!ht]
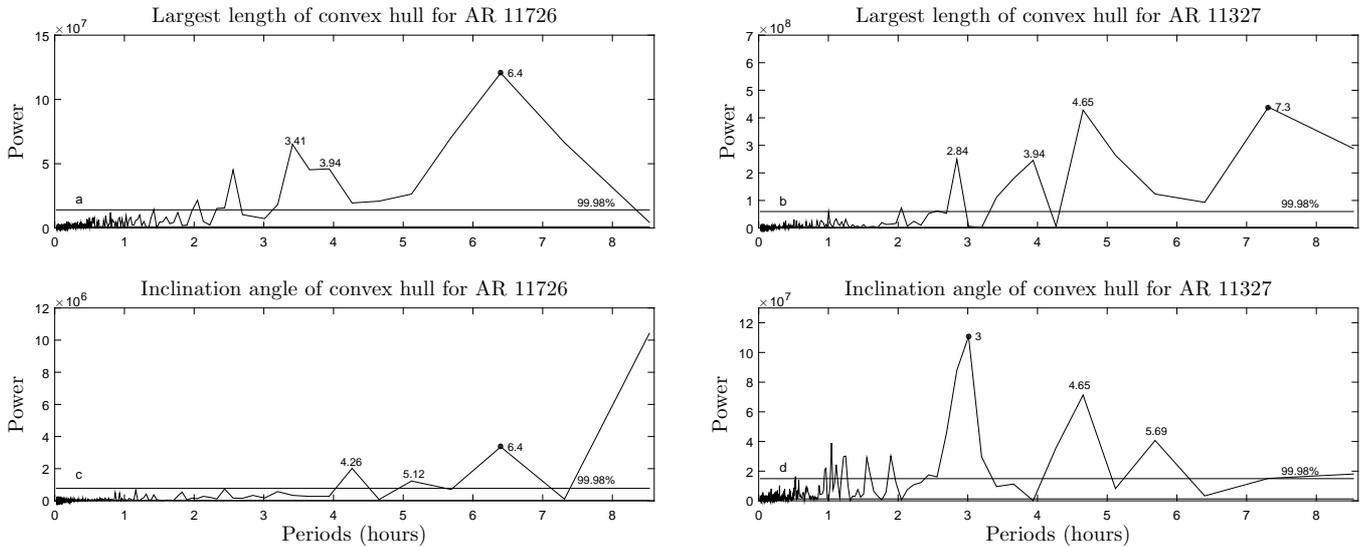

\begin{minipage}[h]{0.5\linewidth}
\center{\includegraphics[width=8.5cm]{fig6.eps}}
\end{minipage}
\begin{minipage}[h]{0.5\linewidth}
\center{\includegraphics[width=8.5cm]{fig8.eps}}
\end{minipage}
\caption{The FFT analysis of the oscillations of the parameters of the convex hull, for ARs 11726  (left panels) and 11327  (right panels), respectively. The vertical line indicates the spectral power. The horizontal axis indicates the periods of oscillation (in hours). Here we removed the highest peaks, corresponding to 12 and 24-hour instrumental periodicities.}
\label{hullperiod}
\end{figure*}

In the convex hull approach, we also performed a Fourier analysis that also showed characteristic periods for the inclination angle and the greatest length between two nonadjacent points of the convex hull. The artificial periods were also removed. After this, we found several periods with a confidence level higher than 99.98$\%$. The spectral analyses of the detrended convex hull data for ARs 11726 and 11327 are shown in Fig.~\ref{hullperiod}. All mean periods and error bars are calculated using the same methods as for the ellipse. All these values with their error bars are also shown in Table~\ref{table}.

\begin{table*}
\caption{Oscillation periods of parameters of the ellipse and the convex hull with error bars. The table represents the characteristic errors using two different methods. Method 1: the half-width of the corresponding power peaks and then taking their mean values. Method 2 is based on finding the standard error of the peaks. All data are given in hours.}
\centering
\begin{tabular}{l c c c c c c c}
 & \multicolumn{3}{c}{AR 11726} & & \multicolumn{3}{c}{AR 11327} \\
 \hline\hline
 &  Periods &  \multicolumn{2}{c}{Errors}  & &  Periods  &  \multicolumn{2}{c}{Errors}   \\
 & mean (min, max) & Meth 1 & Meth 2 & & mean (min, max) & Meth 1 & Meth 2 \\
 \hline
Major axis                      & 4.53  (3.65, 5.69) & 0.39 & 0.49 & & 4.38  (3.2, 5.69)  & 0.32 & 0.59 \\
Minor axis                      & 5.05  (3.65, 6.4)  & 0.54 & 0.65 & & 4.66  (3.65, 5.69) & 0.32 & 0.6  \\
Inclination angle               & 4.56  (3.9, 5.12)  & 0.47 & 0.29 & & 4.7  (3.01, 7.3)   & 0.55 & 0.85 \\
Greatest length (convex hull)    & 4.58  (3.41, 6.4)  & 0.62 & 0.75 & & 4.68  (2.84, 7.3)  & 0.51 & 0.82 \\
Inclination angle (convex hull) & 5.26  (4.26, 6.4)  & 0.6  & 0.51 & & 4.45  (3.01, 5.69) & 0.3  & 0.63 \\
\end{tabular}\label{table}
\end{table*}

\section{Analysis of the oscillations}\label{analysis}
Our aim is to understand the dynamical properties of the ARs as a system of interlinked sunspots and to examine their long-period oscillations. Their dynamic behavior consists of three main components, viz.\ the creation and evolution of the ARs, their translational migration over the solar surface, and their oscillatory motions. The structure of the oscillatory system consists of two main spots of opposite polarity, both anchored in the high-density solar interior, and linked to each other with magnetic loops in the low-density solar atmosphere. These configurations are rather complex because of the structured density profile, and ARs are usually  fragmentized in a set of interlinked sunspots. Strictly speaking, such an oscillatory system can be modeled as a system of interrelated magnetic loops \citep{parker1979}. Therefore, we emphasize here that the notion of the AR depth used throughout this paper is related exactly to this fragmentation of the flux tubes. However, within the scope of the current study we distinguish the {}principal components{} in terms of the two main large spots, thus mimicking the basic system in the form of a single loop. According to \citet{parker1979}, the magnetic flux tube splits into many individual small ones below the solar surface. We assume that at the depth this occurs, the wave is reflected back to the surface. An analytic study of waves and oscillations might be cumbersome from the mathematical point of view (stiff problems). However, here we provide a rather simple interpretation, which enables a first-order estimation of the physical parameters of the ARs. From this perspective, the method can be further developed in the future as a full-scale magneto-seismological tool for ARs with a more rigorous analytical background. We also develop the image moment method, where substructure oscillation spectra will be studied separately.

The ARs consist of two main spots with opposite polarity anchored deep inside the Sun and connected by a loop above the surface. We assume that the long-period oscillations correspond to kink eigenmodes. We consider the characteristic periods of the time evolution of the inclination angle and of the axes length oscillations separately. The inclination angle oscillation is interpreted as the standing second harmonic of the kink-type mode (in our analysis we examine fast and slow kink modes, and conclusions are given below), which implies that the oscillation has a node in the apex of the loop and the AR oscillates as a whole around the axis passing through this node. Our goal is to examine this hypothesis by calculating the characteristic mean propagation speeds (which should be realistic) and, by checking the maximal depth of the oscillation pattern, to provide an estimate of size of the sub-surface (convection zone) part of the AR.
The phase speed of the oscillation mode must be much higher at the solar surface than that at the sub-surface foot points. This description does not rule out a magnetic connection of the sunspots to the base of the convection zone, it merely outlines possible properties of the standing oscillations of ARs. From our analysis we try use the properties of the oscillations to deduce the appropriate depth of the AR anchoring and to which types of mode, fast or slow, the observed oscillations correspond. Using the Fourier periodograms from the analysis above and after excluding the artificial periods, we calculated the characteristic average inclination angle and axes length oscillation periods with a certain error bar (we discussed the methods used for the error estimation above) for both ARs. The obtained values were  4.9 hours and 4.7 hours for AR 11726, and  4.6 hours and 4.6 hours for AR 11327.

To understand the physical properties of the observed oscillations, we need to take the structured environment into account. In addition
to the cylindrical geometry that is usually addressed in this case, following
\citet{Roberts00} and \citet{Edwin83}, we have to take into account that the density profile along the magnetic field lines is rather complex: above the solar surface magnetic loops connect the sunspots, and these loops are embedded in a low-density plasma (with low $\beta$ plasma);~there is an abrupt density transition at the solar surface, and~the density gradually increases with depth (the sub-surface region has high plasma $\beta$).
The development of an MHD formalism for such a complex setup is a rather challenging task, requiring the derivation and solution of multilayer dispersion equations of complex, transcendental type \citep{zaqmur07}. This is far beyond the scope of this study.
As noted by \citet{Roberts00} and \citet{Edwin83}, under the photospheric conditions where $V_{A0}>C_{se}>C_{k}>C_{s0}>V_{Ae}$ (with $V_{A}$ and $C_s$ the Alfv\'en and sound speed at the solar surface (index 0) and in the external medium (index 0)),
the kink speed $C_{k}$ is sub-Alfv\'{e}nic and supersonic ($C_{k}>C_{s0}>V_{Ae}$), while the tube speed $C_T$ (defined below in Eq.~\ref{ctformula1}) is clearly subsonic.
However, this configuration is only appropriate for the surface layer. Because the plasma density increases drastically with depth, the above inequalities are not valid below the surface of the Sun. Furthermore, unless we develop a rigorous theory of such waves (as mentioned above), we have to consider mean values of the quantities following the definition of the average value as
\begin{equation}\label{average}
\overline{f}=\frac{\int_0^h f dr}{h},
\end{equation}
where 0 indicates the solar surface and $h$ a certain height or depth. For pragmatic reasons, we neglected the height of the loops above the surface, that is,\ in the low beta plasma medium, and we assumed that the oscillation of the ARs is predominantly determined by the oscillation of the foot points. In addition, we assumed that the foot points of the sunspots are anchored (fixed) at a certain depth in the convection zone where the modes are reflected, thus enabling a standing wave pattern in the system.
There are several possible causes for such reflections. For instance,~as the density of the plasma gradually increases with increasing depth, the plasma contained in the tube gains more and more inertia and thus the displacement of the tube axis (in case of a kink mode) at some depth becomes significantly smaller than its surface value. Thus, even though the magnetic tube itself may still be directly connected to the dynamo generated toroidal field at the tachocline, the depth of the turning point may not coincide with the total depth of the convection zone. The situation is similar when the reflection of the wave at some depth is due to the structural fragmentation of the flux tube below a certain depth (as is predicted by some sunspot models). In this case, wave propagation occurs only till the depth where the fragmentation starts, and below this depth the morphology of one unified waveguide might split into many smaller branches. In either of these cases, one condition should be satisfied: the actual phase speed of the selected mode (harmonic) must be realistic. We address this quantitative aspect of the problem below. We describe such depths as a kind of turning depths of the modes, which determine the shape of the eigenfunctions related to the modes along the depth and spatio-temporal properties of the modes (wave number, period, etc.). In this description, we assumed that the average values of the characteristic speeds of the medium satisfy the relations
$\overline{C}_{se}>\overline{C}_{s0}>\overline{C}_{k}>\overline{V}_{A0}$ and $\overline{V_{Ae}}=0$. Under these circumstances, the kink speed is determined by the relation of the densities inside and outside of the sunspot and the Alfv\'{e}n speed inside it. So the kink speed reads as
\begin{equation}\label{ckformula1}
 C_{k}=\left(\frac{\rho_{0}}{\rho_{0}+\rho_{e}} v_{A0}^{2} \right)^{\frac{1}{2}},
\end{equation}
and the tube speed as
\begin{equation}\label{ctformula1}
 C_{T0}=\frac{c_{s0} v_{A0}}{\left( c_{s0}^2+v_{A0} ^2\right)^{\frac{1}{2}}},
\end{equation}
assuming that $C_{Te}=0$ (as $V_{Ae}=0$).

It should be also noticed that under the considered regimes, both the kink and tube speed are on the order of the Alfv\'{e}n speed. From this moment, without going into detail of the equilibrium state or of perturbation properties of that equilibrium, we consider the average values of these quantities and build our discussion based on two assumptions, viz.\ ~the turning point (depth of fixation of the sunspot foot point) is situated at the place where the phase speed of the considered oscillation becomes dismissingly small compared to its surface value (in case of 10 000 km it reaches 0.01, for 20 000 km 0.09 and for 40 000 km 0.3 of the amplitude surface value); and~the phase speed is distributed over the depth as an exponential function:
\begin{equation}\label{exponent}
    V_{ph}=V_{ph}(0)\exp \left ( \frac{V_{ph}(0)}{\overline{V}_{ph}} \frac{z}{h}\right ),
\end{equation}
where the depth of the sunspot should be justified based on physical grounds:
\begin{equation}\label{diminish}
    \frac{V_{ph}(h)}{V_{ph}(0)} \ll 1,
\end{equation}
where $V_{ph}(0)$ is the phase speed at the solar surface. We estimated it by calculating the surface value of the Alfv\'{e}n speed. As we show below, we numerically validated the relevance of standing modes with the given wavelength for the cases of fast and slow kink modes separately. We calculated $v_{A0}(0)=B_{0}/\sqrt{4\pi\rho_{0}(0)}$. We inferred the magnetic field strength from HMI magnetograms and detected the mean magnetic field of each AR separately. In this way, we obtained the field mean values $956\pm10$ Gauss and $923\pm10$ Gauss for ARs 11726 and 11327, respectively. The plasma density was taken to be $\rho_{e}=2.2\times10^{-4} kg/m^{3}$. In addition, the estimates of the surface value of the phase velocity for fast and slow kink modes resulting from expressions (\ref{ckformula1})-(\ref{ctformula1}) are shown in Table~\ref{table2}.
We calculated the error estimates for the kink and flute modes using the Alfv\'{e}n speed at the solar surface
$v_{A0}\pm\Delta v_{A0}=\Delta B/\sqrt{4\pi\rho_0}$, where $\Delta B=10\;$Gauss is the magnetic field measurement error in SDO/HMI data. For the fast kink mode, the error of the phase speed at the solar surface was estimated using Eq.~(\ref{ckformula1}), yielding
\begin{equation}\label{dfast}
 \Delta v_{fast-kink}=\left(\frac{\rho_{0}}{\rho_{0}+\rho_{e}} \right)^{\frac{1}{2}}\Delta v_{A0},
\end{equation}
and for the slow kink mode using Eq.~(\ref{ctformula1}):
\begin{equation}\label{dslow}
 \Delta v_{slow-kink}= \left(\frac{C_s}{(C_s^2+v_{A0}^2)^{\frac{1}{2}}}-\frac{C_sv_{A0}}{2(C_s^2+v_{A0}^2)^{\frac{3}{2}}}\right)\Delta v_{A0}.
\end{equation}

In case when the observed oscillations of the inclination angle are interpreted as the first harmonic of the kink mode (i.e.\ $n=2$), with the wavelength $\lambda=2L/n$, while here $L\approx 2h$, and
\begin{equation}\label{averagephase}
    \overline{V}_{ph}=\frac{\lambda}{P},
\end{equation}
where $P$ is the observed period of the oscillations. Next, we calculated the average phase velocity from the solar surface to the footpoints of the system. We concentrated on the kink modes to estimate the mean Alfv\'{e}n speed and its distribution over the depth along the sunspots in both ARs. For slow kink modes, we assumed that the temperature inside the tube is $T_0=3/5 T_e$. To evaluate the sound speed at the solar surface inside the sunspot, we used the photospheric value of the temperature outside the spot, $T_e=5700\;$K.

\begin{table}
\caption{Phase speeds at the solar surface of fast and slow kink modes. All velocity data are given in km/s. The errors are estimated using Eqs.~(\ref{dfast}) and (\ref{dslow}), accordingly.}
\centering
\begin{tabular}{ l c c c c }
 & \multicolumn{4}{c}{Observed phase speeds at solar surface (km/s)} \\
  \hline\hline
 & \multicolumn{2}{c}{Fast kink mode} & \multicolumn{2}{c}{Slow kink mode} \\
 \hline
 & AR 11726 & AR 11327 & AR 11726 & AR 11327 \\
 $\rho_0=\rho_e$   & $4.1\pm0.042$ & $3.9\pm0.042$ & $4.4\pm0.044$ & $4.3\pm0.044$  \\
 $\rho_0=\rho_e/2$ & $4.7\pm0.049$ & $4.5\pm0.049$ & $5.2\pm0.053$ & $5.2\pm0.053$  \\
 $\rho_0=\rho_e/4$ & $5.1\pm0.054$ & $4.9\pm0.054$ & $5.9\pm0.06$  & $5.8\pm0.06$  \\
 $\rho_0=\rho_e/6$ & $5.3\pm0.056$ & $5.1\pm0.056$ & $6.2\pm0.063$ & $6.1\pm0.063$  \\
 \hline
\end{tabular}\label{table2}
\end{table}

The resulting characteristic velocities of the wave propagation are shown in Table~\ref{table3}. To estimate the error of the mean phase velocity values for standing (fast and slow) kink modes, we used the expression:
\begin{equation}\label{dkink}
  \Delta \overline{v}=\frac{L}{P}\frac{\Delta P}{P},
\end{equation}
where $\Delta P$ and $\Delta \overline{v}$ denote the value of the inclination angle and axes length oscillation period errors for estimating kink and flute mode period errors, respectively.
We performed these estimations for the two methods shown in Table~\ref{table}, and the phase speed errors obtained from method~2 are shown in brackets in Table~\ref{table3}.

\begin{table*}[!ht]
\caption{Average observed phase speeds of kink ($v_{ph}=L/P_{inc}$) and flute ($v_{ph}=L/P_{axis}$) modes corresponding to the curves shown in Figs.~\ref{fastkinkmodes} and \ref{slowkinkmodes}. In the table the error bars are calculated with two different methods for the oscillation period error estimation (see Table~\ref{table}). We present both of them, and the error values calculated by method 2 are given within brackets. All velocity data are given in km/s. The errors are estimated using Eq.~(\ref{dkink}).}
\centering
\begin{tabular}{ l c c }
  \hline\hline
  \multicolumn{3}{c}{Mean (min, max) phase speeds (km/s) $\pm$ error meth 1(meth 2) for AR 11726} \\
 \hline
Turning point depth (km) & 10 000 & 20 000 \\
Kink mode                & 1.1 (0.96,1.4) $\pm$ 0.12 (0.09) & 2.3 (1.9,2.7) $\pm$ 0.24 (0.18)   \\
Flute (ballooning) mode  & 1.2 (0.9,1.6) $\pm$ 0.13 (0.16)  & 2.35 (1.8,3.1) $\pm$ 0.26 (0.31)  \\
 \hline
Turning point depth (km) & 40 000 & 200 000 \\
Kink mode                &  4.5 (3.9,5.4) $\pm$ 0.49 (0.37)  & 22.7 (19.3,27.2) $\pm$ 2.45 (1.85) \\
Flute (ballooning) mode  &  4.75 (3.6,6.3) $\pm$ 0.52 (0.63) & 23.7 (17.9,31.5) $\pm$ 2.62 (3.17) \\
 \hline
\\
  \hline\hline
  \multicolumn{3}{c}{Mean (min, max) phase speeds (km/s)$\pm$ error meth 1(meth 2) for AR 11327} \\
 \hline
Turning point depth (km) & 10 000 & 20 000  \\
Kink mode                & 1.2 (0.85,1.8) $\pm$ 0.11(0.19) & 2.4 (1.7,3.7) $\pm$ 0.22 (0.39)  \\
Flute (ballooning) mode  & 1.2 (0.85,1.8) $\pm$ 0.09(0.17) & 2.4 (1.7,3.5) $\pm$ 0.2 (0.35)   \\
 \hline
Turning point depth (km) & 40 000 & 200 000 \\
Kink mode                & 4.8 (3.4,7.4) $\pm$ 0.44 (0.78) & 24.2 (17.1,36.9) $\pm$ 2.21 (3.89) \\
Flute (ballooning) mode  & 4.85 (3.4,7.1) $\pm$ 0.4 (0.7)  & 24.2 (17.1,35.5) $\pm$ 1.99 (3.52) \\
 \hline
\end{tabular}\label{table3}
\end{table*}

\begin{figure*}[!ht]
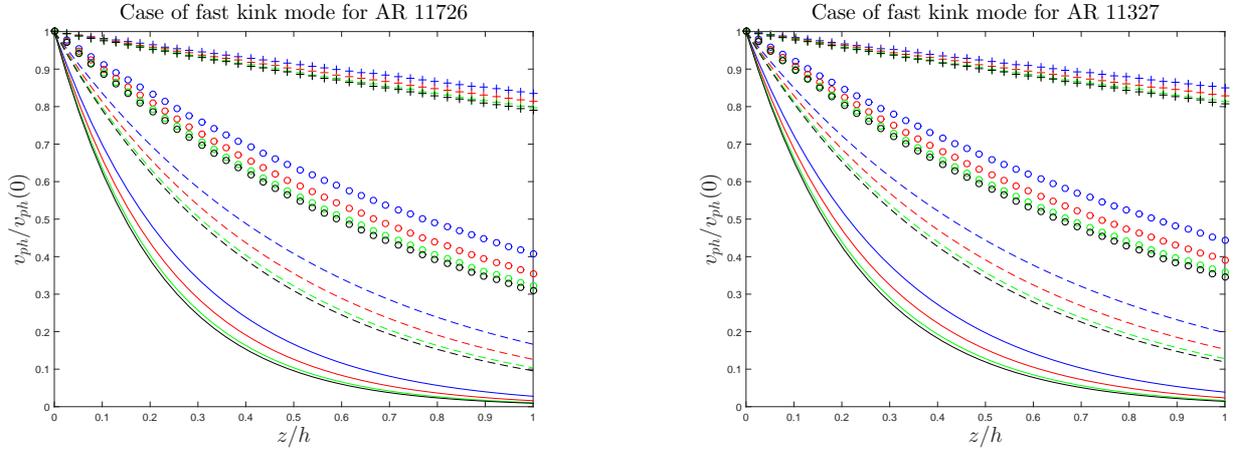

\begin{minipage}[h]{0.49\linewidth}
\center{\includegraphics[scale=0.4]{FKM1.eps}}
\end{minipage}
\begin{minipage}[h]{0.49\linewidth}
\center{\includegraphics[scale=0.4]{FKM2.eps}}
\end{minipage}
\caption{The normalized phase speed Eq.~(\ref{exponent}) $V_{ph}/V_{ph}(0)$ vs.\ the dimensionless depth $z/h$ for the fast kink modes for AR 11726 (left panel) and AR 11327 (right panel). In either of these cases, we consider four options for the surface value of the plasma mass density: $\rho_{0}=\rho_{e}$ (blue curves), $\rho_{0}=\rho_e/2$ (red curves), $\rho_{0}=\rho_e/4$ (green curves) and $\rho_{0}=\rho_e/6$ (black curves), respectively. In addition, it should be noted that we here show curves for four different depth of the wave turning point viz.\ $h=10\;000\;$km (solid line curves), $h=20\;000\;$km (dashed line curves), $h=40\;000\;$km (dotted-dashed line curves) and $h=200\;000\;$km (dotted line curves).}
\label{fastkinkmodes}
\end{figure*}

\begin{figure*}[!ht]
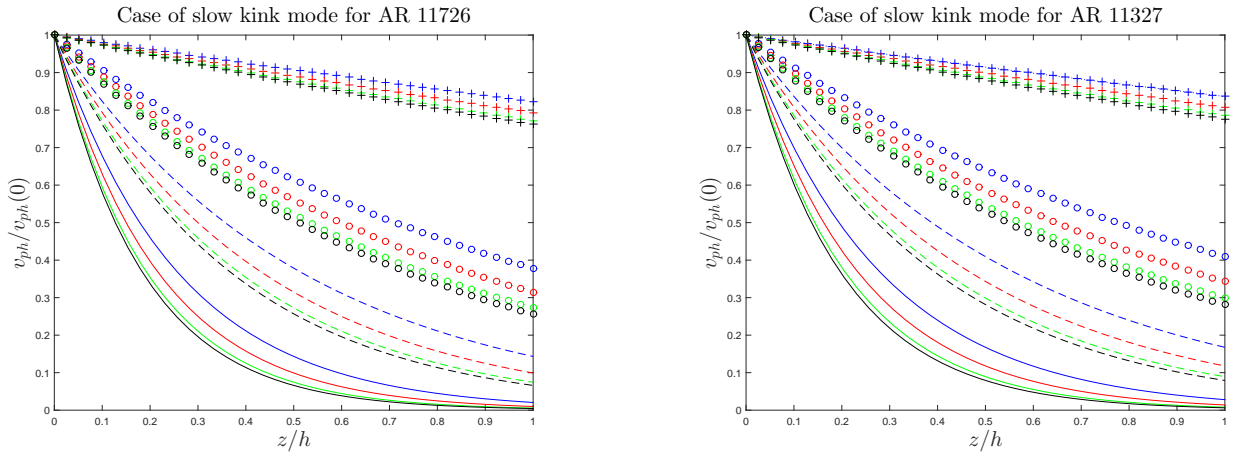

\begin{minipage}[h]{0.49\linewidth}
\center{\includegraphics[scale=0.4]{SKM1.eps}}
\end{minipage}
\begin{minipage}[h]{0.49\linewidth}
\center{\includegraphics[scale=0.4]{SKM2.eps}}
\end{minipage}
\caption{Same quantities as in Fig.~\ref{fastkinkmodes}, but for the case of slow kink modes. For this case, we assume that the temperature inside the tube $T_0=3/5 T_e$. To evaluate the sound speed the solar surface inside the sunspot we use the photospheric value of the temperature outside the spot $T_e=5700\;$K.}
\label{slowkinkmodes}
\end{figure*}

Figures~\ref{fastkinkmodes} and~\ref{slowkinkmodes} show that the obtained depth values are in the range $10\;000\;$km to $200\;000\;$km. The latter depth value corresponds to the thickness of the convection zone, while the other values are justified by direct helioseismic measurements of the sunspot depth \citep{Gizon09}.
Expression (\ref{exponent}) is obtained by taking into account the fact that in the convective zone the plasma density increases gradually with depth and, consequently, the phase speeds of both modes considered should be vanishing at the turning point (compared to its value on the solar surface). Obviously, the values of the mean velocities corresponding to the depth of the convection zone $h=200\;000\;$km (dotted curves in the figures) are inconsistent with this assumption, while the depth of the turning points in the figures converged at about $40\;000\;$km. The analysis we carried out has the potential of becoming a seismological tool for determining the actual depths where ARs are anchored. This tool, however, needs to be further developed in the near future with enhanced mathematical rigor.

There are several arguments to consider when explaining the oscillations of the major or minor axes lengths: (i)~The fundamental harmonic of the sausage mode seems intuitively to be the best candidate. However, this is problematic because as is known, these types of modes are leaky unless the flux tube is sufficiently thick and dense so that the plasma density inside the loop is much higher than outside. The oscillatory system considered here does not satisfy this condition. Moreover, even if it were possible, the characteristic period of the mode has to be twice the period of the kink oscillation, while we found that the characteristic periods of all parameters of the ellipse are similar. In principle, the global sausage mode can be excited, but we cannot resolve the correspondingly twice longer period with our spectral resolution, as addressed above. (ii)~The global vertical kink mode could also explain the {}breathing{} of the ellipse. However, again the global harmonic of the kink mode should have twice the period of the first harmonic, which is not observed in our case. (iii)~Therefore, the best candidate to describe the oscillations of the ellipse axes lengths seems to be the first harmonic of the $m=2$ flute mode. The panels corresponding to the large (top) and small (middle) axes of the ARs in Fig.~\ref{elldetrend} clearly show that they oscillate in antiphase. This pleads in favor of the $m=2$ mode. We calculated the phase speed of the presumed first harmonic of the standing flute (ballooning) oscillation modes, which we associate with the ellipse or convex hull axes lengths (or AR area) oscillations by $v_{ph}=L/P$, where $L\approx 2h$ is the loop length, and $P$ denotes the oscillation period of the axes. The phase speed estimates were made for mean, minimum, and maximum oscillation periods. Similar estimates were made for the kink mode, and the data corresponding to the two types of modes are compiled in Table~\ref{table3}. Therefore, we summarize that the characteristic parameters of the possible standing flute modes also completely agree with the conclusions drawn above.

\section{Conclusions}\label{secconcl}
Using a novel method of automated AR pattern detection, we developed an algorithm that enables the investigation of AR dynamics, based on a least-squares fitting of elliptic patterns. We used the so-called convex hull as a reference for the validation. Using this approach, we identified the AR boundary points by determining magnetic field gradients between neighboring pixels that exceed the threshold value 40$\;$Gauss/pixel. Furthermore, the computational heuristics include mapping of the AR shape on the elliptical shape using a least-squares fitting procedure or a convex hull form. This novel approach enables the determination of the AR dynamics and represents a powerful new diagnostic tool for studying intrinsic kinematic properties of ARs.

We applied the method to the two ARs numbered 11327 and 11726. We discovered that the inclination angles of the corresponding ellipse major or convex hull axes oscillate with respect to\ the solar equatorial plane with characteristic periods of about 4.6 to 4.9 hours. We interpreted these oscillations in terms of the standing second harmonic of kink modes, sustained in two parallel flux tubes (sunspots) with similar properties but opposite polarity, connected by small loops located in the solar atmosphere. We assumed that the ARs oscillate as a whole system, and the corresponding kink mode has a node in the apex above the solar surface. This assumption enabled us to detect the possible distribution of their phase speed values along the tubes in a rather {}seismological{} manner.

We compared the results obtained for fast and slow kink modes, and for each case we concluded the following.
\begin{enumerate}
  \item The mean velocity values corresponding to the entire depth of the convection zone $h=200\,000\;$km  are inconsistent with this assumption as the obtained mean phase velocity values are higher than their surface values. Therefore, our analysis shows that in our setup, the modes propagating to that depth are practically ruled out.
  \item The characteristic depths of the turning point of waves that might satisfy the requirements of the current preliminary modeling must be at most about $40\,000\;$km, where presumably the sunspot structure should break up into smaller flux tubes anchored down in the tachocline, which is in coincidence with helioseismic detection of sunspot depth \citep{Gizon09}, although both methods of depth estimation are rather indicative and require further development of the modeling with stricter mathematical rigor.
  \item Our modeling might become a basis for further development of a seismological tool for the determination of AR structures.
\end{enumerate}

Based on the AR pattern automated detection model, more rigorous magneto-seismological tools can be introduced \citep{Nakariakov01, chorley11, Hoeksema14, Zhang10}, which would include consistent forward- and inverse-modeling components. The latter also concerns the dynamics of both major and minor axes sizes that also show
strong  oscillatory behavior. Here, we showed an illustrative example to interpret these long-period oscillations. The formalism we used is  based on the theory of MHD waves in magnetic cylinders \citep{Roberts00, Edwin83, Nakariakov03}, although the structures we considered have a variable density along the field lines. However, our approach still enables a rough estimation of the mean propagation speeds, and the precision of these measurements is good enough to characterize
the phase speed profiles inside the sunspots semi-qualitatively (quantitatively) and to evaluate the anchoring depth of the ARs. A more thorough analysis of these oscillations requires more sophisticated analytical and numerical methods (tools). This type of investigation, however, is beyond the main scope of the current paper and will be addressed elsewhere.

\begin{acknowledgements}
The work was supported by European FP7-PEOPLE-2010-IRSES-269299 project - SOLSPANET. The work was also supported by Shota Rustaveli National Science Foundation grant DI/14/6-310/12. B.M.S. and M.K acknowledge the support by the Austrian Fonds zur Foerderung der Wissenschaftlichen Forschung within the project P25640-N27 and T.Z. - under project FWF26181-N27. We are grateful to the anonymous referee for the constructive comments on our manuscript that led to significant improvements.
\end{acknowledgements}
\appendix
\section{Least-squares mapping of second-order curves}
A second-order curve that satisfies an algebraic equation of the second degree

\begin{equation}
 Ax^2+2Bxy+Cy^2+2Dx+2Ey+F=0.\end{equation}
Here A, B, C, D, E, and F are the unknowns, (x, y) are the obtained coordinates of the automatically detected data. Appropriately, we obtain a homogeneous system with six unknowns and many equations that always have trivial resolution. We divide all coefficients on $A$ to reduce the number of unknowns by one in the obtained algebraic equations. We can develop a least-squares minimization using these equations, which is optimal among all other methods for the determination. In this way, we obtain formulae for the
coefficients of the second-order curve.

The coefficients, which ensure that the obtained curve is ellipse, must satisfy the conditions: $\Delta\neq0$, $J>0$ and $\Delta/I<0$, where

\begin{equation}
 \begin{array}{ c | c c c |}
 \multirow{3}{*}{$\Delta=$} & A & B & D \\
 & B & C & E \\
 & D & E & F \\
\end{array},
\end{equation}
\begin{equation}
 \begin{array}{c | c c |}
 \multirow{2}{*}{J=} & A & B \\
 & B & C \\
\end{array},
\end{equation}
\begin{equation}
 I=A+C.
\end{equation}

By the boundary points outlined, the second-order curve, which is the ellipse, has been constructed by the obtained coefficients. The parameters of the ellipse have also been found through them, such as the major and minor semi-axes

\begin{equation}\label{majoraxis}
a=\sqrt{ { 2(AE^2+CD^2+FB^2-2BDE-ACF) \over (B^2-AC)( \sqrt{(A-C)^2+4B^2}-(A+C))}},
\end{equation}
\begin{equation}\label{minoraxis}
b=\sqrt{ { 2(AE^2+CD^2+FB^2-2BDE-ACF) \over (B^2-AC)(-\sqrt{(A-C)^2+4B^2}-(A+C))}},
\end{equation}
coordinates of center

\begin{equation}
\begin{array}{c}
x_0={ (CD-BE) \over (B^2-AC)}, \\
y_0={ (AE-BD) \over (B^2-AC)},
\end{array}
\end{equation}
and the inclination angle of the ellipse major semi-axis to the
equatorial plane

\begin{equation}\label{angles}
\varphi=\left\{
\begin{array}{l r}
0, & B=0, A<C \\
\frac{1}{2}\pi, & B=0, A>C \\
\frac{1}{2}cot^{-1}(\frac{A-C}{2B}), & B\neq0, A<C \\
\frac{\pi}{2}+\frac{1}{2}cot^{-1}(\frac{A-C}{2B}), & B\neq0, A>C
\end{array}.
\right.
\end{equation}
\bibliography{mybib}
\end{document}